# A network paradigm for very high capacity mobile and fixed telecommunications ecosystem sustainable evolution

*Francesco Vatalaro*,[§*] *Gianfranco Ciccarella*[†]

§ Dipartimento di Ingegneria dell'Impresa "Mario Lucertini", Università di Roma Tor Vergata, ITALY

† Independent consultant, L'Aquila, ITALY

* Contact person: vatalaro@uniroma2.it

5 June 2020

**Abstract –** The main objective for very high capacity (VHC) fixed and mobile networks is improving end-user QoE, i.e., meeting the KPIs – throughput, download time, round trip time, and video delay – required by the applications. KPIs depend on the end-to-end connection between the server and the end-user device. Not only Telco operators must provide the quality of the needed applications, but also they must address economic sustainability objectives for VHC networks. Today, both goals are often not met, mainly due to the push to increase the access networks bit-rate without considering the end-to-end applications KPIs. This paper's main contribution deals with the definition of a VHC network deployment framework able to address performance and cost issues. We show that three are the interventions on which it is necessary to focus: *i)* the reduction of bit-rate through video compression, *ii)* the reduction of packet loss rate through artificial intelligence algorithms for lines stabilization, and *iii)* the reduction of latency (i.e., the round-trip time) with edge-cloud computing and content delivery platforms, including transparent caching. The concerted and properly phased action of these three measures can allow a Telco to get out of the Ultra Broad Band access network "trap" as defined in the paper. We propose to work on the end-to-end optimization of the bandwidth utilization ratio (i.e., the ratio between the throughput and the bit-rate that any application can use). It leads to better performance experienced by the end-user, enables new business models and revenue streams, and provides a sustainable cost for the Telco operators. To make such a perspective more precise, the case of MoVAR (Mobile Virtual and Augmented Reality), one of the most challenging future services, is finally described.

## 1. Introduction

For some ten years, Europe first established the DAE (Digital Agenda for Europe) industrial policy for the Member States to implement Ultra BroadBand (UBB) objectives, and then promoted the so-called European Gigabit Society (EGS) through the development of VHC (Very High Capacity) networks. VHC pillars for 2025 are two: a dense Europe-wide fabric of optical fiber (or equivalent) access and the development of the mobile 5th generation (5G).

Recently, the EGS objective, as the European Commission (EC) put it forward, started raising concerns about sustainability and credibility in the assumed time frame. According to the European Investment Bank, the required investments to reach DAE/EGS targets account to € 384 bn until 2025 (median scenario), while the market can deliver around € 130 bn, i.e., only one-third of the required amount. Therefore, they concluded that:

*i)* "Gigabits Society targets for VHC networks are far beyond what market forces can deliver";

*ii)* the "investment gap needs to be fulfilled with a substantial degree of public support"; and

*iii)* there is a "high risk of failing to meet Gigabit Society goals" [1].

Therefore, the European mobile and fixed network ecosystem are going to face serious sustainability problems, and the recent COVID-19 outbreak (Sars-CoV-2 coronavirus pandemic) can only make things worse. On March 5th, 2020, the UK Parliament launched a public inquiry on the EGS objectives to ascertain "how realistic the ambition is, what is needed to achieve it, and what the Government's target will mean for businesses and consumers" [2]. According to these authors, other governments in Europe should review their chosen policies towards the EGS rapidly to adjust them to the present scenario. This review is now more urgent than before, as the pandemic consequences are making the Telco industry face the short-medium term challenge to anticipate network improvements as much as possible through a more affordable approach in the light of the new societal scenario and the economic crisis.

When carefully analyzed, the EC's approach appears controversial both from a technical and an economic point of view, as well as for industrial policy strategy. The apparent consequences of the lack of optimality in the broad picture vision designed by the EC for the VHC networks could be more dangerous in the new socio-economic scenario in front of us. The EU Electronic Communication sector is facing, in the best-case scenario, a 2-year delay in the growth. According to Analysys Mason, "COVID-19 will lead telecoms revenue to decline by 3.4% in developed markets in 2020". Their study, published April 2020, points out that "overall revenue declines are expected to amount to 3.4% in 2020 (against a previous forecast of an increase of 0.7%) with a modest rebound of 0.8% in 2021". [3]

It is even more certain that Europe needs to improve the telecommunications infrastructures to cope with the service challenges both present and future and to provide citizens with a better quality of service at an affordable cost since now. However, promoting too far-fetched objectives could backfire and, while risking to not delivering the intended outcomes, they may severely put under stress the European telecom industry as a whole, and delay infrastructures especially in suburban and rural areas.

This paper aims to show an alternative conceptual approach to achieving the objectives set at the mid-decade target date, illustrating a different way of framing the industrial strategy that is less risky, more gradual, and, indeed, able to materialize the EGS targets earlier. The proposed evolutionary approach applies to fixed networks in their migration from copper to full optical fiber access (i.e., FTTH or fiber to the home). It equally well applies to the path from present mobile 3G/4G to the 5G.

For 5G, the Total Cost of Ownership TCO (Capex+Opex) for mobile radio access networks (RANs) is expected to increase sharply. An analysis published by McKinsey & Co. [4] for one European Country, assumed three mobile operators to follow a conservative 5G investment approach, the "TCO for RAN would increase significantly in the period from 2020 through 2025, compared to expected 2018 level. In a scenario that assumes 25% annual data growth, TCO would rise by about 60%". Therefore, in a typical scenario, mobile Telcos will need to develop new strategies for 5G to afford increased network cost vis-a-vis saturated revenues, especially in the over-regulated European Countries. Standard measures will involve cost-saving efforts, but Telcos are also exploring alternative approaches, such as network sharing. With both spectrum sharing and active network sharing, estimates for RAN network savings range between 25% and 40% (CAPEX) and between 20% and 30% (OPEX). At the same time, some limited additional savings are also expected for the mobile backhaul section [5]. Despite that, the development of 5G may be disappointing, unless a mobile Telco develops new profitable revenue models that are still uncertain (e.g., autonomous cars, smart cities). Moreover, it can easily fall under attack from very agile competitors without any infrastructure and acting on top of the network layer.

In several environments, it is difficult for fixed networks to make business plans affordable with an abrupt strategy to replace mixed copper-fiber (i.e., FTTC) networks with FTTH (e.g., see suburban and rural areas).

We organize the rest of the paper as follows. Section 2 introduces the main key applications and network performance indicators (KPIs) that we need to consider to provide UBB services. In particular, live video services are among the most demanding and are limited by the overall network performance. Then, it discusses Quality of Experience (QoE) vs. Quality of Service (QoS) and shows the limitations of the traditional approach to customer satisfaction. In Section 3, we analyze the performance of applications and introduce a simple condition to avoid the risk of being 'trapped' in the inefficient UBB domain. Then, we show the main approaches to address these issues. In Section 4, we show a conceptual procedure to improve network performance progressively and to reduce cost. Then, we show in one very demanding service example, Mobile Virtual and Augmented Reality (MoVAR), how we can achieve network performance targets. Section 5 provides our conclusions.

## 2. Performance indicators and perceived quality

### 2.1 Application services and network services KPIs

Telecommunications networks, both mobile and fixed (and convergent), offer two types of services: application services (or, more simply, applications) and network services:

- *Application services* are the end-users' applications (e.g., web browsing, e-mail, messaging, video streaming, cloud, gaming, 360° augmented/virtual reality). Typically, to provide such applications a software runs on the servers of service providers (e.g., WhatsApp, Dropbox, Skype, Netflix) and on devices used by end-users themselves (e.g., personal computers, smartphones, tablets, TV sets). Today, mostly OTs offer application services and, in a limited way, other Content Delivery Providers (CDPs). In contrast, Telco operators provide a tiny percentage of application services (some 5%, or so).

- Network services are the IP packet transport services that must provide the performance levels required by applications.

Application services mentioned above show a wide variety of applications that do not impose the same IP traffic burden on the networks. Table 1 shows the 2017-2022 projections of consumer traffic types on the internet according to the classification provided by the Cisco VNI [6]. This traffic is consumer IP traffic passing through the internet (i.e., it not related to the network of one single

Telco/ISP). In the following, we will generically refer to Telcos. However, most of what we say applies to both Telcos and ISPs.

Video streaming over the internet requires a share of bandwidth that Cisco VNI expects to grow over 82% of all consumer internet traffic in 2022. Network traffic analyses confirm that we can already estimate HTTP traffic at about 75% today. Moreover, different application services require different quality or performance levels. As an example, the video streaming standard definition (SD) has a much lower resolution than UHD (4K), and then the applications KPIs will be much less demanding.

Edge-cloud computing [7-10] represents an opportunity for a Telco to improve application services performance and to reduce TCO for the provision of network services. The opportunity to provide application services to the end-user from a nearby server, possibly within the access network, without the application traffic having to go through the metro network and the core network, improves the performance and may allow the reduction of the investment needed to ensure KPIs compliance with a given quality level.

To clarify, let us take two simple examples. First, an edge-cloud solution for the Dropbox service could reduce the amount of data crossing the Telco networks by up to 90% [11]. However, we should consider that this data service in 2022 represents only a small fraction of the total consumer traffic, and the expected reduction of overall traffic is no higher than 10%. Moreover, Dropbox contents distribution in the edge-cloud has a high storage cost, as low-cost solutions (e.g., transparent caching) cannot be used due to the small number of end-users that download the same content. Second, with state of the art technology, we can estimate the cut in video traffic at around 50%, but the reduction is related to more than 80% of the total traffic. Then the impact on consumer traffic would be 40%, which is far higher than the expected savings on the data traffic as mentioned above. Movies (VoD streaming) and live events (live streaming) generate most of the video traffic. We can use transparent caching for the effective edge-cloud distribution of their contents.

As a first consequence, it is in the interest of Telco operators to deploy as soon as possible edge-cloud computing technologies to reduce video traffic in the core and aggregation sections of their networks: we believe that the killer application of the ECC is content distribution. Once Telcos deploy edge-cloud computing close to the end-users (in the fixed access PoP and mobile BTS aggregation sites), other services can be distributed, including enterprise edge-cloud, internet of things, gaming, and augmented/virtual reality. Moreover, new architectures such as cloud RAN or virtual RAN can be deployed.

Therefore, to analyze how to improve the performance of applications and to evaluate the advantages of edge-cloud computing, it is necessary to focus on the applications KPIs related to the "end-to-end" connection between the servers running them and the devices presenting them to the end-users [12-14]. The KPIs depend on the Telco networks (access network, metro network, core network) and on the other networks, which signal flows must cross to reach the end-user from the distant server that provides the service.

The main KPIs related to the performance of all the applications and, in particular, of the video services, are:

- *Throughput* (TH), the "speed" at which end-user devices and servers exchange application data. It is one of the most important indicators and can reach up to 1 Gbit/s for applications such as 360° MoVAR, and in the future, it will possibly increase. Throughput is always lower than the bit-rate (BR), the "speed" in the communications channel between end-user and server, due to the congestion control algorithms. For current UBB networks, this difference is very large. The BR and, when the BR is not the bottleneck, the 'distance' metric between the application and the end-user limit the throughput.

- *Latency*, measured through the round-trip time (RTT). Latency must be very low (i.e., in the order of milliseconds) to improve throughput and meet real-time requirements of some services (e.g., telesurgery, autonomous driving, tactile internet). In some cases, applications may require RTT values as low as 1 ms or shorter.

- *Download time*, measuring the response time to end-user requests (e.g., web server response time is the time it takes to display a web page).

- *Video delay*, a live streaming indicator that measures the time, amounting in seconds or several ten seconds, between the instant the camera captures a video frame, and the instant the end-user device displays it on the screen.

The applications KPIs are mainly managed by the Layer 4 (end-to-end applications transport) of the IP protocol stack, and also depend on the network KPIs, i.e., on the transport of the IP packets between the server and the end-user device (end-to-end packets transport). Layers from 1 (physical) to 3 (network) of the IP protocol stack manage the transport of the IP packets. The basic end-to-end network KPIs are:

- ✓ bit-rate, a function of the capacity of the network links. Packets flow between the server and the end-user across the network links. The available bit-rate for one application reduces if the number of flows increases because applications share the bit-rate (fairness principle of the applications transport protocol);

- ✓ the packet loss;

- ✓ the latency measured by the round-trip time that is both one of the main network and applications KPIs.

Below we deal with the relationship between KPIs and different quality definitions of interest to the Telco and the end-user.

## 2.2 QoE and QoS

A rather broad definition of Quality of Experience, provided by ITU-T P.10-2016, consists of "the degree of delight or annoyance of the user of an application or service. It results from the fulfillment of his or her expectations with respect to the utility and/or enjoyment of the application or service in the light of the user's personality and current state." Although not always numerically quantifiable, QoE is the most significant factor in assessing customer experience. Historically, Telcos attempt to derive Quality of Experience from Quality of Service, which intends measuring service parameters objectively through averaging some main technical network KPIs, such as latency, packet loss rate, and bit-rate.

According to ITU-T P.10/G.100, QoS is the "totality of characteristics of a telecommunications service that bear on its ability to satisfy stated and implied needs of the user of the service."

When compared, the two definitions show a substantial difference. On the one hand, QoS is network-centric and provides an "ensemble view" – i.e., it is based on average values – not related to the service experience perceived by a given customer. Its goal is to support network management and improve the average network quality. In general, QoS cannot appropriately meet the expectations of customers, who demand peak-hour rather than average time quality. On the other hand, QoE focuses on performance (or quality) of applications, both technical and subjective. It also has a subjective (but not arbitrary) set of measures from the user's point of view of the overall quality of the service provided, aiming at capturing specific needs. Therefore, QoE is user-centric, and we evaluate it through applications KPIs. In other words, rather than focusing only on technical parameters, QoE provides an assessment of human being expectations, feelings, perceptions, cognition, and satisfaction with a particular application or service.

Some of today's subjective KPIs for the QoE are the abandonment of a given application, complaint calls to a service provider's control center, churn, the amount of time spent with the application, the use of "like," and so on. The applications' subjective KPIs are strongly related to the technical KPIs, e.g., for e-commerce, the conversion rate increases when the catalog download time decreases. Predictable subjective indicators for the future include facial expressions, verbal criticism, and comparison with the past or other applications. When considering technical KPIs, such as throughput in the QoE framework, we intend them in the peak-hour, and their values are closely related to the user's satisfaction. The throughput and the other technical parameters defined above, when evaluated for each customer one by one, are essential technical QoE parameters for video services. The bit-rate is undoubtedly important because it could limit the throughput. However, it cannot be analyzed 'alone' without considering the total throughput required by all active applications, and then it is not a QoE indicator.

## 2.3 The limitation of the traditional approach to customer satisfaction

To improve application performance, a Telco typically works on the transport of IP packets. It implements QoS-based traffic management techniques such as bandwidth reservation and packet prioritization at Layer 2 and Layer 3 of the IP protocol stack. Although these actions may assist in restoring network failures and congested network paths, they are not effective in improving the application performance, because the congestion control limits applications KPIs mainly managed at Layer 4. Then QoS techniques have limited effects towards a better service experience. QoS can help to eliminate through bandwidth reservation the bottlenecks for the available bit-rate.

While QoE mainly focuses on the applications services and QoS on the network services, technical QoE and QoS are related. Traditionally, a Telco tries predicting QoE based on a construct of QoS-QoE models [15]. However, the Telco's predicted QoE turns out to be a loose function of the set of network performance (averaged) KPIs. To make things even worse, the complexity of the underlying interactions often masks the QoS-QoE relationship, as application QoE mainly depends on Layer 4 behavior, as well as on human and context factors – not included in a QoS-QoE model. Consequently, a Telco is generally unable to capture the quality perceived by its end-users one by one.

On the one hand, OTTs and CDPs provide most of the applications delivered by software executed in the servers and the end-user devices. In the following of the paper, we will generically refer to OTTs. However, most of what we say applies to both OTTs and CDPs.

To provide good or excellent QoE, that enables higher revenues both from advertising and from service fees, the OTTs locate their servers close to the Telcos networks and in some cases, when allowed, into the core networks. Telcos do not have direct access to the end-user devices, nor can acquire application information by DPI (deep packet inspection), due to encryption possibly applied by OTTs. Even if DPI worked for some of the OTT services, human and contextual factors would continue to be missing. Moreover, we cannot integrate into service delivery more complex, context-aware QoE models for specific applications.

On the other hand, the OTT has access to end-user devices information by having its application running into these devices. The OTT can then monitor the essential applications technical KPIs (e.g., throughput, latency, download time and video delay), other KPIs (e.g., stall events and playback buffer occupation, that depend on the main technical KPIs). It can also obtain an understanding of the context in which applications are used, such as the user's location (e.g., acquired by GPS sensor and accelerometer), the type of device, and the type of internet connection. The OTT is also familiar with user profiles. The end-user downloads an application to use the service. For some services, each end-user must have a personal account to subscribe to the OTT service. Within the user's account, information is contained, such as the price (if any), service preferences, and in some cases, the end-user characteristics, which may be useful to implement a better forecast of the perceived QoE.

With the information on applications KPIs in hand, the OTT can adapt in near real-time the service quality (such as the video resolution) to the network conditions by adaptive streaming algorithms and can then improve applications KPIs. However, the OTT may still face problems in the client device or the Telco network. These problems limit the OTT in always providing a good QoE to the end-users.

The OTT usually has no QoE issues in the big internet thanks to the distribution of applications and content from their centralized data centers (such as hyper-scale data centers) to medium or small data centers located at the edge of the big internet, as close as possible to and interconnected with the Telco's networks. Today, Content Delivery Networks (CDNs) massively manage the QoE. About twenty years ago, the OTTs started to use content delivery (CD) platforms located within CDNs. Therefore, they have an established technology and can ensure the quality level required by the different applications on the big internet.

CDN's whole premise is to distribute contents and applications as close as possible to the end-user. In principle, we can also use this approach within the Telco network. However, the Telco network architecture today does not provide IP Layer 3 user plane visibility between the core network and the end-user equipment. Therefore, CD platforms can only be located in the core network and are less effective as the 'distance' to reach the end-user is large.

Modern video streaming platforms, possibly located in a CDN, adopt appropriate application layer adaptive technologies, such as the DASH (Dynamic Adaptive Streaming over HTTP) MPEG standard (ISO/IEC 23009-1) [16], to allow adapting as much as possible to network conditions to comply to end-to-end technical QoE requirements individually for each end-user (Figure 1).

The OTT (or the CDN provider) does not directly control the network service. So, when network conditions degrade due to RTT and packet loss rate increase or to available bit-rate reduction, it can only reduce the source quality level to adjust end-to-end throughput as requested by feedback signals directly coming from the media playback controller.

Therefore, DASH continuously adapts the short-term throughput to provide the best possible streaming quality. This adaptation implies variable video presentation quality level that, however, is much better than uncontrolled data speed random throttling, which causes very unpleasant video stalling conditions. However, as forthcoming contents and applications require high or very high throughput (e.g., for 4K streaming quality and 360° augmented/virtual reality), this approach alone becomes less effective, and we cannot avoid looking better into the Telco's network.

To improve the QoE in their networks, Telcos should adopt the same approach used by OTTs: this asks for placing content delivery platforms close to the end-user premises. Acting on IP Layer 4 functionalities provides a performance improvement. It requires the distribution of some core functions – such as the EPC (Evolved Packet Core) in 4G mobile networks, and the BNG (Broadband Network Gateway) in fixed access networks. The goal of the EPC and BNG distribution is providing user plane IP Layer 3 visibility that, as said above, the Telco network architecture presently does not provide, to enable the distribution of contents and applications close to end-users. This distribution of contents and applications entails complexity higher than that of the well-established CDN solution on the big internet.

### 3. Throughput vs. bit-rate: the UBB domain

Let us now discuss what causes applications KPI degradation and how we can overcome it. We mostly concentrate on analyzing the throughput, which is the most important KPI, with some mention of the other main KPIs.

On the one hand, when customers think of internet "speed," they are implicitly referring to the performance of the applications (i.e., to the end-to-end throughput). On the other hand, "speed" for Regulators and Telcos today is generally the bit-rate (i.e., the speed of the transmission channel) of the fixed and mobile access networks. Moreover, the speed test measurements provide the bit-rate (to be more precise a proxy of the bit-rate) in the network segment between the end-user device and an intermediate server that makes the measurement. This server is different from the OTT server that executes the applications and then provides the services to the end-users. This misunderstanding is a source of confusion, as very often people believe that a network is 'fast' if the access bit-rate is 'fast' while not considering that the most important KPI for a fast network is the application throughput, that depends on the end-to-end connection between the server and the end-user device and not only on the access network bit-rate. What is worse, the confusion pushes towards increasing the access infrastructure cost for a potentially poor end-user performance experience [17, 18].

In a given geographical area, the network throughput – e.g., the throughput handled by one single access POP that aggregates and manages the traffic of all the access links for that network area – must be able to ensure the performance required by all the simultaneously active application services.

The throughput of one single application, in the given network area, is limited by the minimum between two values: the available bit-rate, which depends on the type and the number of simultaneously active applications, and the maximum network throughput that is limited, due to congestion control, by round-trip time and packet loss ratio. By considering one single active TCP (the transport layer protocol adopted more than 95% of times) application, the following simplified but accurate expression provides its throughput [19]:

$$TH \leq min\left\{ c\, \frac{MSS}{RTT}\, \frac{1}{\sqrt{PLR}}, BR \right\} \qquad (1)$$

where $c$ is a constant (typical values are 0.9÷1.2), $MSS$ is the maximum TCP segment size (typical 1460 Byte for video services), $RTT$ is the round-trip time, $PLR$ is the packet-loss ratio, and $BR$ is the available bit-rate (i.e., the bit-rate the application can use). According to eq. (1), the product $RTT \cdot \sqrt{PLR}$ is the server to end-user 'distance' metric. In general, RTT and PLR depend on network technology, its topology, and total traffic. Eq. (1) gives the upper bound of the steady-state application throughput for deterministic packet loss for one TCP connection and traditional end-to-end "windows" based congestion control mechanism. Eq. (1) is valid for any traditional TCP implementations (e.g., Tahoe, Reno, CUBIC) with different values for the constant, $c$. Table 2 shows some throughput values vs. $RTT$ and $PLR$, assuming $BR$ is not the bottleneck.

Heuristically, we can also assume eq. (1) to provide the boundary between the UBB bit-rates region and the BB/NB (broadband/narrowband) bit-rates region in a legacy network. In fact, for some given $RTT$ and $PLR$ values, in BB/NB networks, the bit-rate is 'low' (i.e., $BR$ is the bottleneck for $TH$) and the so-called bandwidth utilization ratio $r = TH/BR$ is 1 (or close to 1). On the contrary, in UBB networks the $BR$ is 'high', $RTT$ and $PLR$ limit the throughput, then $TH < BR$ and, as $BR$ grows, $r$ can even become very small ($r << 1$): under this condition, bit-rate resources are wasted in one or more part(s) of the network.

Therefore, the bandwidth utilization ratio can significantly degrade in UBB networks as compared to legacy networks. Since the bit-rate increase is often deployed mainly in the access network, this is the part of the network that wastes bit-rate resources. Therefore, focusing on the bit-rate increase in the access network, that is presently the only objective of Regulators and Telcos, conflicts with throughput limitations given by $RTT$ or $PLR$, or both, and by the available bit-rate between the big internet and the access network. As an example, for LTE networks, the measured cumulative distribution function of the radio-link bandwidth utilization ratio, $F(r)$, was given in [20] for "large" TCP downlink flows (i.e., > 5 s and > 1 Mbyte) not concurrent with other streams. On average, the bandwidth utilization ratio is $r_a =$ 34.6 %, and the median ratio is only $r_m = 19.8$ %. The median ratio shows that one single application uses the radio channel bit-rate for about one-fifth of the time. Then, the radio access bit-rate is far from being the bottleneck for the application.

In the segment from the big internet to the access network, we get the UBB network throughput by aggregating the traffic of all the access links managed by one access POP. Then, the bandwidth utilization ratio is, in general, much higher than the bandwidth utilization ratio in the access network. Therefore, in some cases, *BR* in this network segment is the bottleneck for the performance of the applications (Figure 2). This figure provides the average peak-hour bit-rate per line in Italy's fixed networks (aggregation, metro, and core networks) measured in 2015-2017 and published by AGCOM [21]. The peak-hour bandwidth increase year over year in this network section is 25-30% (the former value for xDSL accesses and the latter value for FTTx accesses). Extrapolating the linear increase to 2020 overall *BR* is about 2 Mbps. This result evidences the large gap between the *BR* available for applications in the access network and the network from the big internet to the access [18].

To improve the application throughput and the bandwidth utilization ratio, when the available bit-rate is not the bottleneck, *RTT* and *PLR* must be reduced. We can obtain this result by the distribution of contents and applications close to the end-users and by appropriate physical layer improvements.

Note that a low value for the bandwidth utilization ratio, *r*, harms Telco's economic sustainability. It increases network costs and does not allow network monetization, mainly based on network performance. When throughput service requirements become more severe – as it is with time, when going from SD to HD and further on (see Table 3) – retaining the legacy network architecture causes the Telco falling deeper and deeper into the "UBB access trap": by investing in access technologies only, the network costs increase, and the customer satisfaction tends to be very poor. Therefore, we define the UBB access bit-rate limit condition, which we should not violate, through the following relationship:

$$BR \gtrsim \sum_{i=1}^{n} \frac{MSS_i}{RTT_i} \frac{1}{\sqrt{PLR_i}} \qquad (2)$$

derived from eq. (1) with $c = 1$, where now $BR$ is the cumulative bit-rate required by all $n$ applications that are simultaneously using the access link. In general, $n$ is small for residential applications (e.g., less than 3-4).

The end-to-end bandwidth utilization ratio for a given geographical area must consider both the access network links and the network segment between the big internet and the access POP. The latter network segment today generally has a much higher utilization ratio ($r \approx 1$ in the access-to-big internet network segment), because, in most geographical areas, the peak hour available average bit-rate for any active end-user is much lower than the access links bit-rate. The UBB cost inefficiencies are then related to the access network's very low throughput over bit-rate ratio ($r \ll 1$).

We can provide the throughput required by the applications and can reduce the network cost acting through rate compression techniques. If a given value of bit-rate is not the bottleneck, by decreasing $PLR$ and $RTT$, we can increase the throughput and improve the bit-rate utilization.

Let us now separately consider throughput reduction by compression techniques (case 1), $PLR$ reduction by access links stabilization (case 2), and $RTT$ reduction by edge-cloud computing (case 3).

### 3.1 Reducing throughput through video compression (case 1)

Data compression works on the content before transmitting it over the network. In UBB networks, the video traffic will be more than 80% at the peak hour. In order to distribute very high quality video content, we cannot neglect new video compression techniques that strongly reduce the source throughput.

Modern lossy video compression techniques reduce the throughput needed for an application and, therefore, the bit-rate needed to achieve this throughput, if the bottleneck is not due to $RTT$ or $PLR$, or both. In such conditions, while the end-to-end bandwidth utilization ratio, r, is unchanged, the source throughput reduction lowers the network's cost as it reduces peak throughput, and typically improves application performance. Bandwidth utilization ratio in the access network for one single application is generally worsened.

Therefore, advances in video compression are crucial, and studies are underway on techniques to improve low complexity 2-D video encoding (MPEG-5 standard). Recently, MPEG approved the start of work on the next 'MPEG-5 Part 2' standard based on a solution named PERSEUS Plus [22]. It has a data stream structure defined as two-component streams, a primary stream decodable from a hardware decoder, and an enhancement stream that is suitable for software processing implementation. Experiments in the field compared the 'legacy' channel (h.264) broadcast in SD at 1.8 Mbit/s with the same 'PERSEUS-enabled' channel in HD/720p at 460 kbit/s at least at similar quality. It descends about a four-fold reduction factor, which can bring the video compression ratio to about 1:160.

All this may have a significant impact on the time profile of operators' CAPEX investments, allowing, for example, the fruition of HD content even with access network having lower bit-rate performance than previously. However, the compression technique choice is under the OTT's control rather than Telco's control.

For VoD, compression techniques can reduce the required throughput. However, for live streaming, the compression time could be too long and increase the video delay. Today, VoD 4K streaming requires a minimum throughput of 15 Mbit/s, while live 4K streaming requires 25 Mbit/s to reduce the processing time needed for compression and to avoid video delay increase. Therefore, compression techniques can provide limited throughput reduction for live streaming due to video delay constraints.

## 3.2 Reducing PLR through access lines stabilization (case 2)

Packet loss is caused either by congestion in network routers or by transmission bit errors. According to Ookla's speed test, the PLR of the four main mobile networks in Italy measured in December 2018 resulted between 0.38% and 0.83%. Packets loss can never be null, and it acts as the implicit TCP feedback control signal for many network congestion control algorithms.

In the DSL access, frames corruption is a consequence of bit errors. They produce a loss of quality at the channel level, monitored through the so-called code violations that originate threshold-crossing alarms. A sequence of frame losses is a consequence of line instability, a metric derived from outage times related to synchronization losses, and poor performance of line-initialization monitored through parameters such as, e.g., Failed Full Initialization Count [23]. Line instability is a non-stationary process almost impossible to predict. However, Telcos can modify the DSL profile, on a line or bundle basis, to improve line stability and quality, which manifests through higher bit-rate.

Certainly, optical fibers are inherently more stable. However, the Wi-Fi home distribution degrades quality and is still the dominant factor of line instability. About 70% of today's end-users adopt Wi-Fi for indoor signal distribution. In the Wi-Fi section of a wireline network, the primary degradation factors are radio frequency interference due to nearby signals, signal attenuation due to distance, multipath fading, hardware faults, and home setup misconfigurations.

Line instability is also a problem in wireless networks, both for 4G/5G mobile communications and for fixed wireless access. Some leading causes are interference among cells (e.g., small cells superimposed on one macrocell), and multipath fading.

In general, a fundamental trade-off holds between stability and bit-rate: the higher a line's bit-rate is set, that same line's likelihood of instability increases. Hence, it may be necessary to decrease some unstable lines' bit-rate to ensure stable operation. Other lines, by contrast, may be able to increase their bit-rate and remain acceptably stable. Overall, the number of higher-speed lines increases, and we can monitor this average rate metric to ensure optimized stabilization and bit-rate.

Therefore, the use of algorithms for surveillance and quality control of the Telco's lines (copper, optical fiber, and wireless) can allow the removal of physical layer bottlenecks, which directly affect packet loss, including those generally prevailing due to poor performance of home Wi-Fi.

The two *PLR* components can be considered independent; therefore, we have:

$$PLR = PLR_1 + PLR_2 \tag{3}$$

where $PLR_1$ is the component mainly related to the network segment from the access POPs to the server across the big internet, and $PLR_2$ is the component in the access network, including the home network. $PLR_1 = f\,(RTT)$ and depends on network queues in the IP network. $PLR_2$ depends on physical layer impairments (i.e., bit errors).

Therefore, the approach to PLR containment is twofold:

1) to reduce $PLR_1$, we need to improve the transport infrastructure quality in terms of transmission speed and number of packets/s managed by the routers and the applications transport protocols (Layer 4), to bring content delivery platforms and applications closer to the end-user, or both;

2) to reduce $PLR_2$, we can rely on analytics based on Artificial Intelligence (AI) paradigms, whereby the Telco's access network performance can improve by regularly and automatically tuning its components.

Some advantages of optimizing access networks through AI tools include proactive online response to problems as they arise, better allocation of resources (wired or wireless) to dynamically improve the quality of the customer experience (and the speed), as well as the association of the client's reaction to optimization strategies for better use of the available infrastructure resources.

To the aim of reducing $PLR_2$, measuring the stability level is a crucial factor. However, algorithms in practice only assign one out of a few stability levels to each line: very stable, stable, unstable, and very unstable. These four levels have a very high correlation with the customers' propensity to complain [24].

By doing so, the Telco can use QoE indicators such as (*i*) the trouble ticket rate, and (*ii*) the dispatch rate:

- *Trouble ticket rate*: A Telco's call center usually opens a trouble ticket when a customer complains. Among such calls, some are technical and related to the operation of the network's physical layer. A Telco desires reduction of customer-call volumes. A good measure for end-user QoE is the trouble-ticket rate of technical nature, calculating the monthly percentage of such complaints to the number of lines. The daily ticket rate can be volatile with noticeably different patterns over weekends, so the software tools generally consider a 7-day moving average.

- *Dispatch rate*: Customer complaints that the Telco cannot solve remotely result in a technician dispatch on the field (one of the highest costs for a Telco). The dispatch rate is the percentage of dispatches to the number of lines measured every month. As with the trouble-ticket rate, a 7-day moving average is used.

When aiming at further reducing $PLR_2$, we need to work on the physical layer. Still, we must always balance its reduction with the adverse effects on throughput due to the limitation in packet payload and the increase in $RTT$. Reducing $PLR_2$ may reduce the application throughput: for example, increasing forward error correction reduces the packet data payload, while a negative effect on $RTT$ is due to interleaving.

### 3.3 Improving throughput and other application KPIs by edge-cloud computing (case 3)

We should improve applications KPIs without increasing the network cost, which depends on the peak hour load. When possible, we can reduce the network load by adopting: a) the most efficient video compression techniques, which entails lower applications throughput and, as a consequence, lower bit-rate in the communications channel, and b) optimized access network performance by installing AI-based surveillance and control software providing lower $PLR_2$. However, due to internet traffic exponential growth year after year, and to service requirements becoming more and more stringent, this approach turns out to be insufficient. Therefore, we should deploy edge-cloud computing platforms to achieve lower round trip time and, in many cases, lower $PLR_2$.

Edge-cloud computing works above the network layer, at the transport layer and can improve the applications KPIs presented in Section 2.1. It also operates above compression and $PLR_2$ reduction and can coexist with these techniques to improve KPIs. However, we must carefully consider the interaction among ECC, video compression, and $PLR_2$ reduction to avoid a possible negative impact on the applications KPIs.

In the ideal condition of adopting both video compression, techniques to reduce $PLR_2$ and transport protocols that avoid congestion with new flow control mechanisms (e.g., BBR [25]), for each flow one can aim at $RTT$ and $PLR$ not limiting throughput between the server located far away and the end-user device. Therefore, the bandwidth utilization ratio in the network section between end-user and server $r \approx 1$, and to improve application performance increasing the end-to-end bit-rate would be effective. Consequently, ECC platforms would not produce a throughput increase, and then ECC could be considered unnecessary. However, even in this limit condition, the remaining KPIs (such as video delay and download time) may still be inadequate for an acceptable level of QoE. Furthermore, if we also consider the TCO, we can save on network costs by using ECC, as a result of containment of traffic and the needed total peak throughput in the network segment between the ECC and the big internet. Thus, even under such an ideal condition, ECC may turn out to be necessary.

The transparent cache is one main ECC component capable of achieving traffic containment. It locally stores and continuously updates contents, not altering the end-to-end application logic in a fully transparent way to the content provider and the end-user [26]. While already adopted in CDN nodes located at the Telco's core network or farther away, transparent caching is much more effective when brought closer to the end-user.

From a functional point of view, a transparent cache is a local repository which stores the most popular contents (e,g., the most requested high definition hit movies, the currently most clicked video clips or web pages) supplying such contents whenever a nearby end-user requests them after the content provider completed authentication and authorization. By locally storing copies of the most frequently requested contents, the transparent cache can increase applications throughput and reduce download time and video delay, thus improving all the technical QoE KPIs. However, to work correctly, a transparent cache must be dynamic in selecting the locally most requested contents while updating the memory with new ones as it detects changes in the user behaviors and expectations. Transparent caching also requires visibility of the HTTP address, which many times can be encrypted. For this reason, it requires collaboration between the Telcos and the OTTs. Collaboration models based on transparent caches located on the internet at one Telco network border are well established for some twenty years between OTTs and CDN providers.

Zipf's law provides the theoretical foundations for the advantages of using transparent caching. The relative frequency with which end-users request contents follows a Zipf-like distribution, where the relative probability of a request for the *n*-th most popular content is inversely proportional to n$^\alpha$ with $\alpha$ taking on some value less than unity, typically ranging from 0.64 to 0.83 [27]. Therefore, it is enough to store a limited number of contents in the nearby transparent cache to achieve high values for the probability that cache delivers the content, or hit-ratio, *HR*, so avoiding a high number of requests and contents to traverse the network. For $\alpha = 0.8$, only 10% of content stored provides $HR \approx 50\%$, which means cutting in half the heavy video traffic.

Transparent caching is effective in reducing IP downstream data traffic in the network segment between the application server and the ECC. Then, the network upgrade measures taken to manage the growth of the IP traffic volume have a much lower cost. Consequently, the ECC network architecture TCO is lower than the traditional (legacy) network TCO if the ECC network cost is lower than the network upgrade cost for the segment between the ECC and the application server, a condition often met in practice [13, 14].

Let us assume a three-level network model (Figure 3): core network, edge network – generally comprised of a metro component and an aggregation component – and access network. To fix the ideas, in order to provide a rough evaluation of the throughput advantage, we assume five core nodes (CNs) and 25 metro nodes (MNs), while the total number of access nodes (ANs) is $n_{AN} = 250$. In our architecture, rings connect the MNs to the CNs (the average number of MNs per ring is five). The ANs connect to the MNs via rings, and the average number of access nodes per ring is ten.

Below, we only present throughput performance improvement (reference [14] also provides the cost analysis). We evaluate the total traffic throughput improvement for the *i*-th node through the so-called Speed-Up, *SU*(*i*), defined as follows:

$$SU(i) = \frac{TH_q(i)}{TH(i)} \tag{4}$$

where $TH(i)$ and $TH_q(i)$ in Gbit/s denote the total throughput of applications managed by the $i$-th node without and with an ECC platform, respectively. Conservatively, we assume $PLR$ being the same under both conditions. Therefore, $SU$ is a function of $RTT$ ratios, only (see eq. (1)):

$$SU(i) = f\left(\frac{RTT_q(i)}{RTT(i)}\right) \tag{5}$$

where $RTT(i)$ and $RTT_q(i)$ are the round trip time without and with ECC, respectively. We model both $RTT$ parameters as random variables and consider the average values. The speed-up, $SU(i)$, provided by ECC platforms that utilize transparent cache located in the $i$-th node is given by:

$$SU(i) = HR(i)\left(\frac{RTT(i)}{RTT_q(i)} - 1\right) + 1 \tag{6}$$

where the transparent cache hit-ratio, $HR(i)$, is the cache efficiency, equal to the probability that the content is delivered by the cache when the user plane IP layer is visible, and the HTTP address is not encrypted, or the Telco can decrypt it.

Therefore, the Network Speed-Up, $NSU(i)$, for the ECC platform with the transparent cache located in the $i$–th access nodes, $1 < i < n_{AN}$, is:

$$NSU(i) = \frac{\sum_{j=1}^{i} TH_q(j) + \sum_{j=i+1}^{n_{AN}} TH(j)}{\sum_{j=1}^{n_{AN}} TH(j)} \; . \tag{7}$$

We partition the total downstream network traffic among metro nodes according to the law $Y(i) = a \cdot i^m$ with $m = -0.6$ (Figure 4 (a)). Then, we distribute the traffic of each metro node to the access nodes according to the law $Y'(i) = a' \cdot i^{m'}$ with $m' = -0.99$ (Figure 4 (b)). Then, the fraction of the traffic managed by each access node is the product of the percentage of its metro node with the corresponding percentage of the access node. Parameters $a$ and $a'$ are used to normalize the distributions to 1. These distributions were adopted through a traffic analysis based on peak bandwidth data from Analysis Mason [28] and Cisco VNI [5], and on the distribution of end-user traffic measured in Italy [29].

The location of the ECC platforms in the ANs starts from the node providing the highest cost-saving (as defined in [14]). In Figure 5 we report the speed-up of each single access node ($SU(i)$, dashed lines) and the network speed-up ($NSU(i)$, solid lines), defined as the speed-up related to the total network throughput, and obtained by increasing the number of access nodes where the ECC platform is deployed. We order the subscripts of the access nodes in the figure according to cost savings (i.e., the first node has the highest saving, and the last has the lowest saving). When we locate the ECC platforms in all the access nodes, the NSU has the highest value. We consider two scenarios with different average $RTT$ values between the end-users and the access nodes and between the access nodes and the big internet. Scenario A refers to a lower network $SU = 1.75$, and scenario B refers to a higher network $SU = 3$. The increase in the number of access nodes equipped with the ECC improves network $SU$ and the network application throughput if the network bit-rate is not the bottleneck.

In practice, we can readily obtain $SU$ values in the range of about 2 to 4: if $SU = 2$, the throughput with transparent caches is two times the throughput without transparent caches.

## 4. A cost-saving approach to very high capacity network deployment

VHC networks deployment must ensure the respect of end-to-end applications KPIs, while at the same time cost-effective solutions must be devised for two main network segments: the access network, both fixed and mobile, and the IP network from the access POPs to the big internet interconnection.

Cost-effective ECC architecture can improve end-to-end applications KPIs. Moreover, it can provide cost savings for both the fixed and the mobile networks, from the access to the big internet, due to the contents and application distribution close to the end-users that significantly reduces the peak throughput and for the mobile access network (Radio Access Network).

ECC platforms can provide either network cost saving or cost time displacement for mobile and fixed UBB networks, because they reduce RTT and PLR, so improving the bandwidth utilization ratio. Today $r$ is much different in UBB access networks ($r << 1$) and in the network segment from access to big internet ($r \approx 1$). Therefore, we should think of the end-to-end performance to better equalize bandwidth efficiency and improve the application performance for the end-user. Contrary to this cost-effective approach to get performance improvement, a strong push by Regulators towards improving the fixed access networks bit-rate only, not having regard to the end-user QoE objective, is providing an unbalanced and inefficient outcome. For fixed and, in many cases, mobile access networks, the minimum bit-rate provided to any end-users is generally much higher than the bit-rate today available for the applications in the Telco networks from the access POPs to the big internet interconnection.

## 4.1 Evolutionary approach to access network deployment

As discussed above, in both fixed and mobile access networks (with the possible exception of networks in rural areas), the bit-rate is, in general, much higher than that available between the access network and the big internet. The application performance limit is due to the bit-rate that applications can use in the network segment from the end-users access points to the big internet, or the distance (evaluated in terms of $RTT$ and $PLR$) between the end-user device and the server providing the service. The significant increase in the bit-rate of fixed and mobile access networks, which often requires enormous investment, does not correspond to an equivalent improvement in application performance (due to the limits in the network segments from the access POPs to the big internet interconnection). Moreover, capital expenditures do not provide revenues increase. Some critical issues for the economic sustainability of Telco's business models derive and can also lead to low customer satisfaction.

While the EC is promoting a new objective, the one for the achievement of VHC networks in Europe, the investments approach in access networks should change to allow them to be more gradual. To fix ideas with an idealized scenario, the plot in Figure 6 shows three curves corresponding to a bit-rate of 10 Mbit/s, 100 Mbit/s, and 1 Gbit/s, respectively. Let us suppose that for each of the curves, different fixed access technology is needed: for the rightmost curve, a copper network may be sufficient, for the intermediate one, a copper-fiber hybrid network, and finally for the leftmost one the all-fiber network. If we focus on a brownfield network condition, it is evident then that each curve in the graph corresponds to different levels of investment, more intense if we move from the right curve to the left. We can easily extend the concept to mobile networks with different radio link frequencies, RAN bit rates, and cell densities. It also applies to convergent networks.

Now, let us think of the applications that run in these networks: as we have seen, video streaming applications will reach about 80% of the total traffic in two years. Therefore, let us stick with these applications only. In the evolution of the services that the networks carry on, today, video services are mainly SD, while HD services are beginning to penetrate the market. In the coming years, we can expect video quality to increase, so 4K Ultra HD (UHD) is starting to appear with 15 Mbit/s throughput for VoD and 25 Mbit/s for live streaming. Over time, virtual and augmented reality services will begin to appear for the large masses of consumers. These services, as is the case for video alternatives (SD, HD, UHD), will go through several seasons, with increasing quality demand, which in Figure 6 we point out through four stages: Early Stage (ES); Entry Level (EL); Advanced Experience (AE); Ultimate Experience (UE) [30] (see Table 4).

If we think about the transition from the SD/HD video service of today to the future UE virtual and augmented reality service requiring the highest performance, it means moving from point (A) to point (B) shown in Figure 6. Of course, we can think of many transition paths, but if we want to take into account the gradualness of investments without the client perceiving performance limitations as the more valuable services advance, the dotted arrows in Figure 6 show a possible optimal path. The rapid vertical downhill transitions from one curve to another correspond to a change of technology in the access network, without changing the *RTT* value. Bringing contents and applications in the fixed access POPs closer to end-users can improve *RTT*. Lower *RTT* provides higher throughput and then increases the bandwidth utilization ratio, *r*.

The main issue suggesting the gradual transformation of the fixed access network to FTTB/FTTH is the very low bit-rate utilization in the access network, being the throughput limited by the end-to-end bit-rate and by the distance between the server and the end-user device. The ECC roll-out can be considered the key enabler for VHC networks that will have to ensure the very challenging end-to-end performance of future services such as 360° virtual and augmented reality, not only for fixed services but also the highly demanding mobile ones (MoVAR).

As we can see, a network will not be able to provide the future MoVAR services if the network's latency will not fall below the millisecond, besides ensuring end-to-end Gigabit per second speed per active end-user. Furthermore, if we do not reduce the latency down to such a level, any attempts to enlarge the bandwidth are useless, and the associated investment essentially wasted.

Figure 6 shows how the evolution of Telco networks should ideally follow a path characterized by three time-phased elements: *a)* increase in quality and speed of all the applications and, in particular, video services; *b)* decrease the RTT; *c)* increase the end-to-end bit-rate (not only in the access network); and so on.

Such an evolutionary approach, at least in principle, is optimum for network development. Network investments can follow the demand curve, and the Telco operator can reinvest the revenues in the network enhancement without resorting too much to bank credit and minimizing the investment risk. By doing so, and considering that the amount of money to be invested is limited, digital-divide in a Country can be minimized, as there will be no overspending in the most profitable areas at the expense of the less attractive suburban and rural areas. Therefore, it is also inherently more socially responsible.

Being a "continuum model," obviously, this approach cannot be followed straightforwardly, due to several kinds of practical constraints. However, having it in mind can help avoid severe mistakes in network planning, such as anticipating too early the expensive fiber optics roll-out to homes without taking advantage at the right time of technologies such as AI network surveillance and control, and edge-cloud computing.

## 4.2 An exemplary case: Mobile Virtual and Augmented Reality

One of the most challenging services in the future will be 360° MoVAR. It is difficult to provide this service even with present 5G, as well as with fiber optics terminated to Wi-Fi. Provision of the end-user with the so-called "ultimate experience" will require extremely high guaranteed throughput (in the order of 1 Gbit/s) and latency well beyond present achievable values (i.e., $\leq 1$ ms).

In providing 360° MoVAR key performance indicators are very challenging. In fact, for complete solid angle spatial rendering, the human visual field of view (FOV) needs to be accounted for: horizontal FOV is about 180°, and vertical FOV is 135°. Binocular vision, which is the base for stereopsis and is needed for 3D vision, covers slightly less than 120° of horizontal FOV. Therefore, foveal vision is ±30°, while remaining peripheral vision up to 90° each side is not binocular (only one eye can see). In such an external part, the human brain perceives much the movement, so that it can almost instantly rotate eyes. Displays FOV cannot be too limited, as this could contrast the immersive effect and, even worse, it can induce the syndrome of "simulator sickness." Therefore, displays use now foveated rendering, a kind of graphic rendering with eye tracking integrated into a virtual reality interface that significantly compresses the image quality in the peripheral vision to reduce the rendering workload for about ten times.

All this is the basis for the throughput requirement for MoVAR. A key factor in avoiding the simulator sickness for an extended viewing time (> 15 min) is excellent visual quality. This quality level asks for a significantly higher spatial resolution and a higher frame rate than is currently the case in many virtual reality demonstrations: at least 47,000 pixels seem necessary. Therefore, within foveal vision, 170M pixels are needed, and additional 45M pixels within the peripheral vision. The total number of pixels is 215M (minimum), and with 8 bit/pixel, we get 30 frames/s and a gross data-rate of 48 Gbit/s. With a lossy compression factor between 15 and 30 times, the needed (net) bandwidth is not lower than 0.7÷1.4 Gbit/s. To provide the needed QoE for MoVAR, this is the range of values required for the application throughput.

Therefore, Figure 6 gives a requirement for *RTT* of 1 ms or less. As a final observation, we understand that, for future systems, the bandwidth-latency equation provides an essential requirement stricter than the usual traffic capacity requirement. The ECC's transparent cache will diffusely locate at the mobile system's small cells and, in fixed networks, at any building distributors. As a result, the visibility of the network layer must also move deeper and deeper near the end-user terminal, due to application performance constraints that overcome the customary capacity limitations for networks. We can envisage that future networks will be latency-limited rather than limited by interference or bit-rate.

## 5. Conclusions

The main contribution we presented in this paper is related to the definition of the VHC network deployment framework that can address both performance and cost issues. The main objective for VHC fixed and mobile networks is improving end-user QoE. Towards this aim, we must meet the KPIs – throughput, download time, round trip time, and video delay – required by applications. KPIs depend on the end-to-end connection between the server and the end-user device. For VHC networks, Telcos must provide very high speed and must address economic sustainability objectives.

Today, often, both objectives are not met, mainly due to the push to increase the access networks bit-rate without considering the end-to-end application KPIs. To this aim, Telco networks architecture must change, improve lower layer performance through AI-based algorithms, and deploy ECC platforms that distribute contents and applications close to end-users to provide both performance improvements and network costs savings. The continuous run towards higher and higher access bit-rates, without feedback control on the actual end-to-end applications throughput, is a "UBB trap," which we should avoid as much as possible, as it certainly brings about increased cost while the end-user benefit is uncertain.

In this paper, we proposed working on the end-to-end optimization of the bandwidth utilization ratio, $r$, because it leads to better performance experienced by the end-user, provides a sustainable cost for the Telco operators, and may enable new business models and revenue streams. We showed that there is margin to achieve the EC's VHC network targets, if the definitions of VHC networks are focused on the end-to-end connection, leveraging the edge-cloud computing architecture. This approach is a significant paradigm shift compared to what some policy-maker circles generally maintain.

Three areas of concurrent development are urgent, and already implementable with state-of-the-art technology:

*i)* Reduced video bit-rate through new compression standards;

*ii)* Reduced router-independent packet loss component with the implementation of network surveillance and control algorithms based on AI tools;

*iii)*    Development of edge-cloud computing architecture to jointly improve applications performance and to reduce the network TCO.

These areas are not entirely under Telco operators' control. However, a Regulators' holistic approach could push the market towards what we believe is the right direction, promoting progress in the telecommunication industry and improving the Telcos market positioning.

By better aligning investments and service demand according to the conceptual approach delineated in this paper, Europe may always avail of a network at the QoE level requested by citizens. While Europe will not delay the Gigabit/s throughput needed for the most demanding services, it can level investments not to leave behind the rural areas due to unless unavoidable concentration of scarce economic resources in the urban and other heavy-traffic areas. As an additional benefit, therefore, less burden can be expected for the Member States, as Telco operators, both mobile and fixed, will benefit from more economic resources and an increased willingness to invest due to limited risk.

This paper attempted to organize ideas collected through years of examination of the European policies on telecommunications, and we wrote it while a catastrophic virus epidemic is hitting Europe, and a potentially systemic economic downturn is at the view. If this unexpected crisis has to show something to policy-makers, it is that no Country or European region can wait for fiber deployment everywhere – in fact, they need somewhat higher speeds for capillary provision of services such as work from home and distance learning now and at much lower investment both in terms of money and time.

| Consumer Internet Traffic, 2017–2022 | 2017 | 2018 | 2019 | 2020 | 2021 | 2022 | CAGR 2017–2022 |
|---|---|---|---|---|---|---|---|
| **Internet video** | 56 | 77 | 105 | 140 | 184 | 240 | 34% |
| **Web, email, and data** | 12 | 15 | 19 | 23 | 27 | 31 | 22% |
| **Online gaming** | 1 | 3 | 4 | 7 | 11 | 15 | 59% |
| **File sharing** | 8 | 7 | 7 | 7 | 7 | 7 | - 3% |

**Table 1: Consumer internet traffic by subsegment (2017–2022) in Exabyte per month (1 EB = $10^{18}$ Byte), (Source: Cisco, [6]).**

| PLR (%) | RTT(ms) 0.1 | 0.5 | 1.0 | 1.5 | 2.0 | 2.5 | 3.0 | 3.5 | 4.0 | 4.5 | 5.0 | 5.5 | 6.0 | 6.5 | 7.0 | 7.5 | 8.0 | 8.5 | 9.0 | 9.5 | 10.0 |
|---|---|---|---|---|---|---|---|---|---|---|---|---|---|---|---|---|---|---|---|---|---|
| 0.05 | 10,447 | 1,045 | 522 | 348 | 261 | 209 | 174 | 149 | 131 | 116 | 104 | 95 | 87 | 80 | 75 | 70 | 65 | 61 | 58 | 55 | 52 |
| 0.10 | 7,387 | 739 | 369 | 246 | 185 | 148 | 123 | 106 | 92 | 82 | 74 | 67 | 62 | 57 | 53 | 49 | 46 | 43 | 41 | 39 | 37 |
| 0.15 | 6,032 | 603 | 302 | 201 | 151 | 121 | 101 | 86 | 75 | 67 | 60 | 55 | 50 | 46 | 43 | 40 | 38 | 35 | 34 | 32 | 30 |
| 0.20 | 5,223 | 522 | 261 | 174 | 131 | 104 | 87 | 75 | 65 | 58 | 52 | 47 | 44 | 40 | 37 | 35 | 33 | 31 | 29 | 27 | 26 |
| 0.25 | 4,672 | 467 | 234 | 156 | 117 | 93 | 78 | 67 | 58 | 52 | 47 | 42 | 39 | 36 | 33 | 31 | 29 | 27 | 26 | 25 | 23 |
| 0.30 | 4,265 | 426 | 213 | 142 | 107 | 85 | 71 | 61 | 53 | 47 | 43 | 39 | 36 | 33 | 30 | 28 | 27 | 25 | 24 | 22 | 21 |
| 0.40 | 3,694 | 369 | 185 | 123 | 92 | 74 | 62 | 53 | 46 | 41 | 37 | 34 | 31 | 28 | 26 | 25 | 23 | 22 | 21 | 19 | 18 |
| 0.50 | 3,304 | 330 | 165 | 110 | 83 | 66 | 55 | 47 | 41 | 37 | 33 | 30 | 28 | 25 | 24 | 22 | 21 | 19 | 18 | 17 | 17 |
| 0.60 | 3,016 | 302 | 151 | 101 | 75 | 60 | 50 | 43 | 38 | 34 | 30 | 27 | 25 | 23 | 22 | 20 | 19 | 18 | 17 | 16 | 15 |
| 0.70 | 2,792 | 279 | 140 | 93 | 70 | 56 | 47 | 40 | 35 | 31 | 28 | 25 | 23 | 21 | 20 | 19 | 17 | 16 | 16 | 15 | 14 |
| 0.80 | 2,612 | 261 | 131 | 87 | 65 | 52 | 44 | 37 | 33 | 29 | 26 | 24 | 22 | 20 | 19 | 17 | 16 | 15 | 15 | 14 | 13 |
| 0.90 | 2,462 | 246 | 123 | 82 | 62 | 49 | 41 | 35 | 31 | 27 | 25 | 22 | 21 | 19 | 18 | 16 | 15 | 14 | 14 | 13 | 12 |
| 1.00 | 2,336 | 234 | 117 | 78 | 58 | 47 | 39 | 33 | 29 | 26 | 23 | 21 | 19 | 18 | 17 | 16 | 15 | 14 | 13 | 12 | 12 |

*Throughput (Mbit/s)*

**Table 2: Some throughput values as a function of *RTT* and *PLR*.**

| Video resolution | SD | | HD | | UHD |
|---|---|---|---|---|---|
| **Device** | Smartphone | TV | Smartphone | TV | TV |
| **Netflix** | - | 3,0 | - | 5,0 | 25,0 |
| **YouTube** | 0,5 | 3,0 | 3,0 | 2,5[1]-5,0 7,0-13,0 (live) | 15,0-25,0 |
| **Amazon Prime Video** | - | 0,9 | - | 3,5 | - |
| **Apple TV** | - | 2,5 | - | 8,0 (6,0)[2] | - |
| **DAZN** | 2,0 | - | 3,5 | 6,5-8,0[3] | - |
| **Legend:** **Throughput values expressed in Mbit/s** **Note 1: 720p** **Note 2: 1080p HD  (case Mid Definition a 720p)** **Note 3: High frame rate** | | | | | |

**Table 3: Downstream throughput typical values recommended by some widely used internet video platforms (Source: Platforms published data).**

| Technology | Time forecast (*) | Field of view | Resolution | Pixels | Color depth | Frame rate | Compression ratio | Equiv. TV resol. | Needed Throughput (Mbit/s) | Max time of use |
|---|---|---|---|---|---|---|---|---|---|---|
| **Early** | 2016 (now) | 90° | 2K | 3840x1920 | 8 bit | 30 fps | 165:1 | 240p | 25 | 20 min |
| **Entry level** | 2018 | 90° | 4K | 7680x3840 | 8 bit | 30 fps | 165:1 | SD | 100 | 20 min |
| **Advanced experience** | 2021 | 120° | 8K | 11520x5760 | 10 bit | 60 fps | 215:1 | HD | 400 | 60 min |
| **Ultimate experience** | > 2021 | 120° | 16K | 23040x11520 | 12 bit | 120 fps | 350:1 | UHD | 1.000 | 60 min |
| **(*) Time forecast (2016) looks optimistic at this time of writing: Ultimate experience will most likely appear > 2025, while the intermediate stages are still on the way.** | | | | | | | | | | |

**Table 4: Expected evolution of virtual and augmented reality. (Source: Huawei, [26]).**

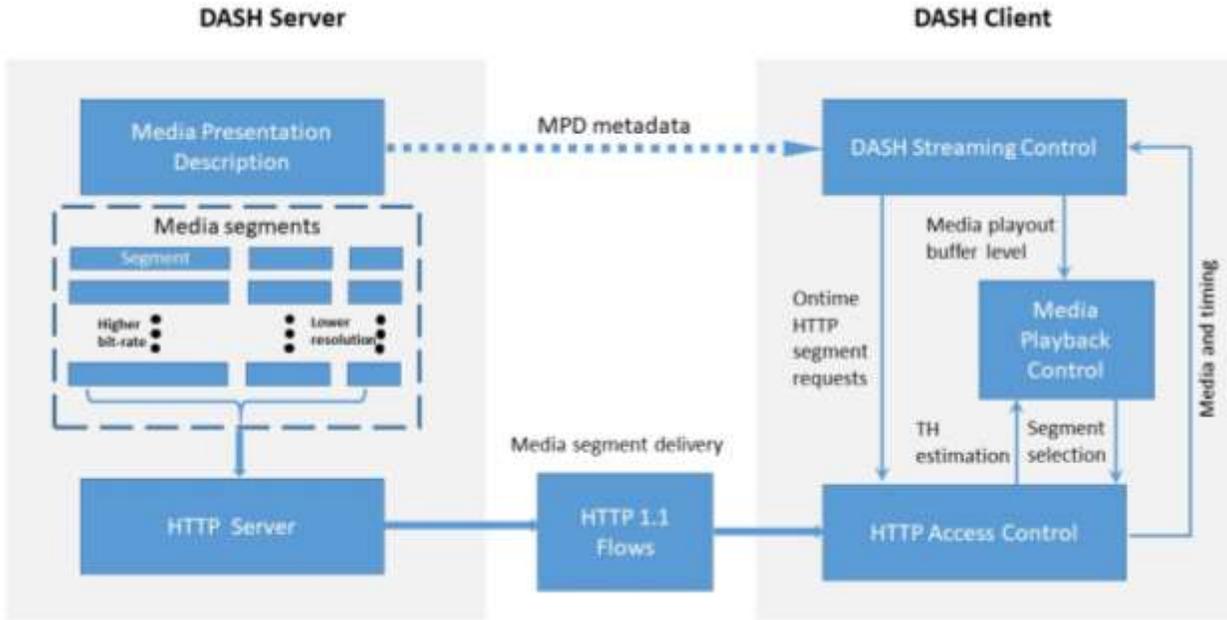

**Figure 1: MPEG-DASH system architecture.**

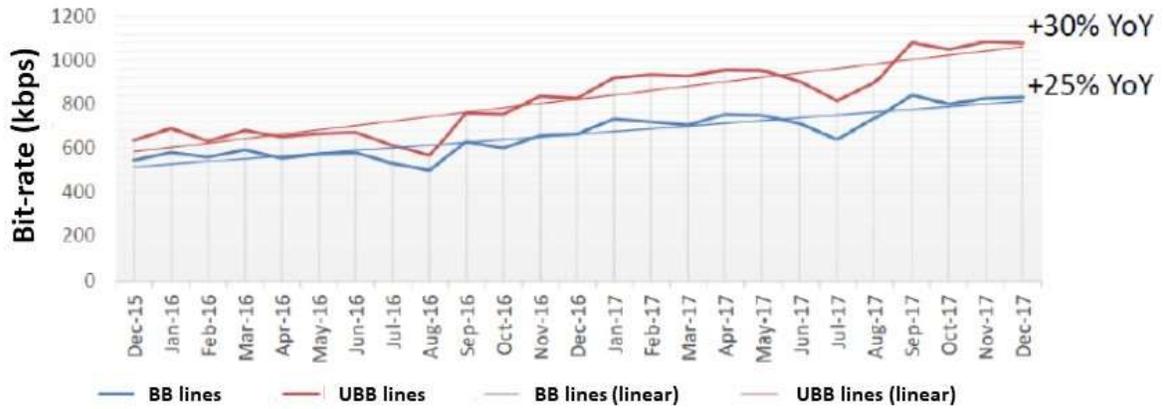

**Figure 2: Average peak bandwidth/total lines in Italian fixed networks measured over two years (Source: Agcom, 2018).**

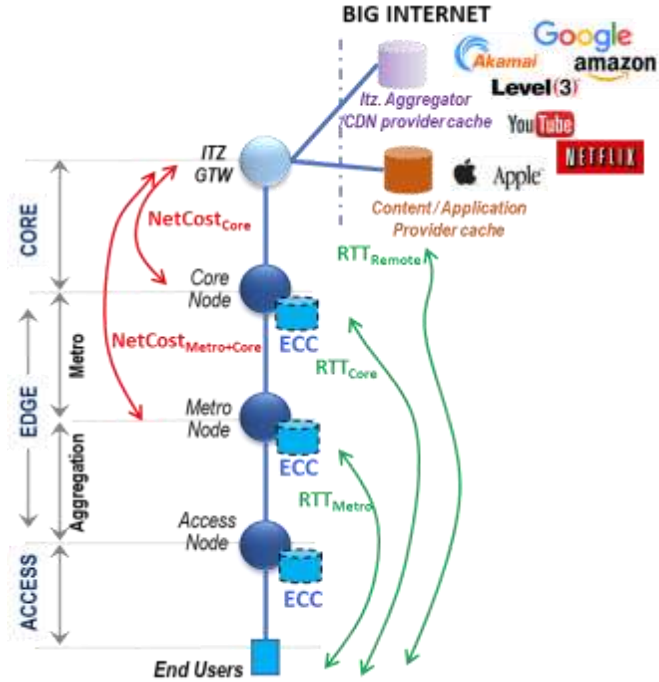

**Figure 3: Reference three-level network architecture and model key parameters.**

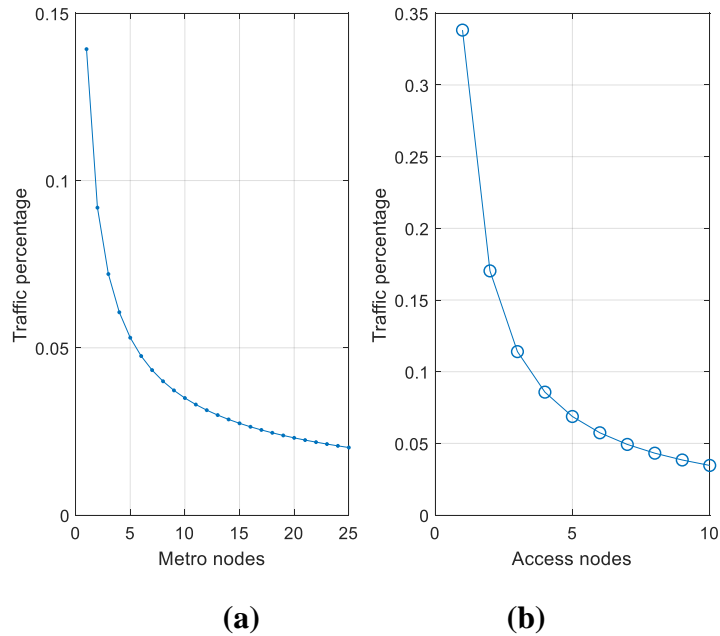

**(a)**                    **(b)**

**Figure 4: Traffic distribution for: (a) metro nodes and (b) each set of access nodes connected to one metro node.**

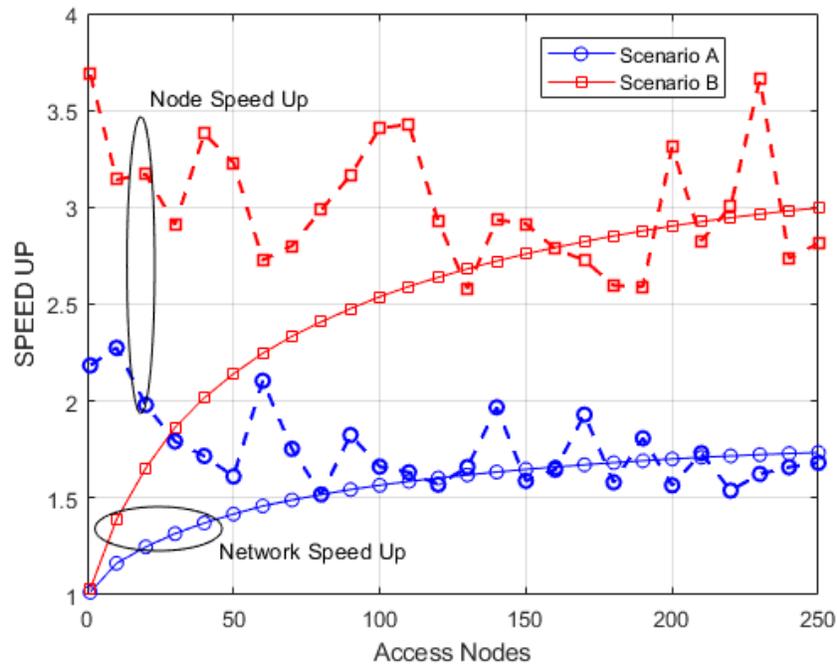

**Figure 5: Network Speed Up vs Access Nodes with ECC for scenario A (NSU=1.75) and scenario B (NSU=3).**

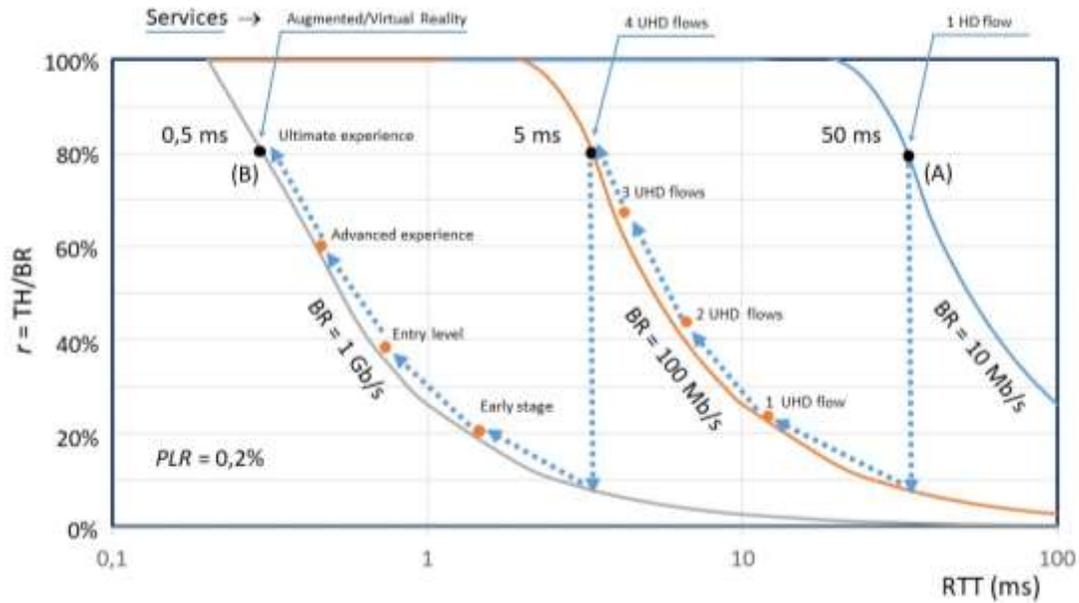

**Figure 6: Conceptual evolution of a network to provide new services while continuously readapting the network infrastructure.**